# Towards the Development of an LLM-Based Methodology for Automated Security Profiling in Compliance with Ukrainian Cybersecurity Regulations

Daniil Shafranskyi[1], Iryna Stopochkina[2], and Mykola Ilin[3]

*[1,2,3] National Technical University of Ukraine*
*"Igor Sikorsky Kyiv Polytechnic Institute", Beresteiskyi Ave, 37, Kyiv, 03056, Ukraine*

**Abstract**
In recent years, the pace of development of information technology in various areas has increased drastically, forcing cybersecurity specialists to constantly review existing processes in order to prevent unauthorized access to confidential information. Using Ukraine as a primary case study, this paper explores the integration of international best practices, specifically ISO/IEC 27001 and the NIST Cybersecurity Framework, into national regulatory systems. A focus is placed on the transition from traditional compliance models to risk-based approaches, exemplified by the recent adoption of the Ukrainian normative documents. Furthermore, we propose a methodology for automating the development of target security profiles using Large Language Models (LLMs) enhanced by Retrieval-Augmented Generation (RAG). By integrating a vector database of national regulations and organizational policies, the proposed RAG-based advisor reduces manual complexity, minimizes human error, and ensures alignment between technical controls and legal requirements.
This study contributes to the field by providing a structured workflow for AI-assisted cybersecurity management in environments characterized by high-intensity hybrid threats.

## Introduction

The rapid development of information technologies has significantly impacted the way organizations store, process, and transmit the data. The increasing reliance on digital infrastructure across government, business, and critical sectors has led to a growth in cybersecurity threats, such as unauthorized access, data breaches, and cyberattacks targeting information systems [1]. To address such challenges, many countries have adopted systematic approaches to information security management based on internationally recognized standards and frameworks. Among the most widely used are ISO/IEC 27001 standard and the collection of NIST publications on information security, which provide structured methodologies for identifying security risks, implementing protective controls, and maintaining robust information security systems [2]. As for now, these two are considered the best in information security management and are widely applied in organizations of different kinds.

**Problem statement.** In Ukraine, the creation and maintenance of information security systems is regulated by a number of legislative and regulatory acts that define the legal and organizational basis for protecting information in information and telecommunications systems. At the same time, the national regulatory framework is increasingly taking into account international standards and practices in order to ensure compatibility with global approaches to cybersecurity and strengthen the national IT infrastructure [2,3].

An important component of modern information security management is the development of security profiles that define specific security requirements and controls for particular information systems. International frameworks, particularly those developed by NIST, provide both methodologies and comprehensive sets of security controls. At the same time, the rapid development of artificial intelligence technologies opens new opportunities for improving the processes of cybersecurity management. According to [4], modern AI-driven defensive techniques, which include anomaly detection, real-time threat analysis and adversarial training, apply machine learning to detect attack patterns, anticipate emerging threats, and optimize security decision-making. Nevertheless, AI can also be applied to automate risk assessment and support the development of target security profiles for complex information systems.

**Research gap.** A significant gap remains between the adoption of Ukraine's new risk-oriented authorization approach and the practical means available for implementing it. Practitioners still lack a clear and reproducible workflow for profile tailoring. This makes the development of an automated, explainable, and regulation-aware method of target security profile (TSP) derivation a relevant research problem.

**The purpose of this paper** is to develop a structured methodology for automated target security profile formation, based on a baseline profile and supported by a RAG-enhanced LLM approach, in compliance with the evolving Ukrainian cybersecurity regulatory framework. The paper aims to decrease existing research gap.

**The main contributions of the work are:**
1. The method for Target Security Profile derivation under the Ukrainian regulatory framework is proposed.
2. An LLM-RAG model of architecture for generating draft control-level decisions with traceable justification and standardized output structure is proposed.
3. An initial experimental validation of the methodology through a comparative assessment of several large language models for different cases with expert supervision is performed.

## 1. Related work and Regulatory Context of Information Security in Ukraine

Ukraine has a number of laws and regulations governing information security within the country. The most important of these are the Laws of Ukraine, which defines key concepts such as information and information security. The regulatory documents of the ND TPI framework, developed by the State Service of Special Communications and Information Protection of Ukraine (SSSCIP), establish a structured catalog of security measures for automated systems. In 2021, the number of ND TPI were updated to better align with the NIST Cybersecurity Framework, which will be discussed in more detail in the next section.

Our attention is concentrated on ND TPI 3.6-006-24 [5]: "Procedure for Selecting Information Protection Measures, the Protection of Which is Mandated by Law and Does Not Constitute State Secret, for Information Systems." Introduced in 2024, this document represents a modern pivot in the regulatory framework. It provides a standardized methodology for selecting security controls for systems handling sensitive but non-classified information, specifically aligning these choices with the organization's risk profile.

This document introduces the concepts of different types of security profiles (Fig.1) and their role in shaping a resilient security strategy:
1. *Base Security Profile*: A minimum mandatory set of security controls established for a specific system category such as Low, Moderate, or High impact. It serves as the universal foundation and the mandatory starting point for any security design or architecture [5].
2. *Sector-Specific Security Profile*: An adaptation of the Base Profile tailored to the specific needs of an industry such as Finance or Healthcare or a technology like Cloud or IoT. It standardizes security requirements within a sector and optimizes costs by focusing on threats and technologies relevant to that specific field [5].
3. *Target and Adapted Security Profiles*: The final, customized set of controls designed for a specific information system, derived from either a Base or Sector-Specific Profile. These are necessary for practical implementation, ensuring security [5].

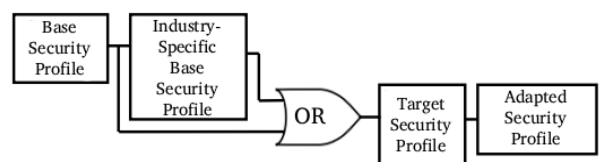

**Figure 1**: Types of security profiles and their interrelationships (from [5])

Among the documents that form the basis of Ukraine's regulatory framework for information security, NIST SP 800-53 ([6]) and ISO/IEC 27001 ([7]) are the most commonly cited.

According to the analysis [8], NIST SP 800-53 is a robust catalog of controls that cover various areas of security, which was originally designed for the needs of the US government, but quickly gained popularity among private companies, while ISO/IEC 27001 is an international information security standard that details the core processes a company needs to practically manage its cybersecurity risks.

The approach to security profiles incorporated by Ukraine is based precisely on NIST 800-53, and, as stated in [9], ND TPI 3.6-006-24 (Procedure for Selecting Information Protection Measures, the Protection of Which is Mandated by Law and Does Not Constitute State Secret, for Information Systems) essentially incorporates security measures from NIST SP 800-53 Rev. 5 and translates them.

The process of developing security profiles is traditionally performed by security experts who analyze regulatory requirements, identify potential threats, and select appropriate security controls from control catalogs. However, the growing complexity of modern information systems and the increasing volume of regulatory documentation make this process time-consuming and difficult to perform manually.

The author of the publication [10] discusses the use of artificial intelligence in the financial sector for processing and structuring transactional data, reports. The Retrieval Augmented Generation (RAG) technology proposed as a solution is a tool for addressing given tasks, as it combines retrieval-based and generative models to achieve greater accuracy and efficiency. However, the author primarily suggests using RAG to generate context-aware compliance reports or regulatory filings, which does not entirely align with the objectives of this article.

Article [11] discusses a system consisting of several RAG agents designed to automate the verification of regulatory compliance for software requirements. Within the proposed framework, two independent AI agents act as Compliance Checkers, analyzing requirements against a policy document, while another separate agent serves as a Report Analyzer, which compares the reports provided by the Compliance Checkers, locates inconsistencies, and instructs the models to review their analyses, after which it generates a final report. While this architecture has a number of advantages, such as documented improvements in accuracy and simplified scalability, it may not be suitable for assisting in the creation of security profiles in Ukraine. First of all, using a larger number of models significantly increases costs in terms of both money and resources, making this method financially unfeasible for small businesses. Secondly, a multi-vendor approach significantly increases the likelihood of a supply chain attack, since each individual vendor is a potential link through which data leaks or unauthorized access could occur. This factor can prove critical for government agencies or any other companies that handle restricted-access information. A viable alternative could be using different models from the same vendor (for example, OpenAI's GPT-5.4 and GPT-5.4-mini, the former for more complex tasks, the latter, a lighter one, for generating follow-up questions and performing basic operations), or simply sticking to a single-model architecture in cases where the improvement in the quality of generated responses does not justify the higher price.

In paper [12] the RAG approach was successfully implemented for a complex problem of binary vulnerability analysis, that shows its prospectivity for precise tasks.

A significant advancement in structured knowledge representation is the Extended Semantic Networking (ESN) framework proposed in [13]. Unlike conventional RAG systems that typically operate by retrieving and injecting textual snippets, ESN introduces a novel architecture for constructing dynamic semantic networks. The framework functions by expanding an initial LLM-generated conceptual graph with concept-relation pairs extracted from external documents, resulting in a hybrid and evolving knowledge graph. However, for real-time compliance automation, such approach could hinder the system's responsiveness compared to more streamlined RAG implementations.

The authors of [14] focus on developing a multifactorial risk assessment methodology, integrating international standards (ISO/IEC 27033, NIST SP 800-53) and the Zero-Trust concept into modern wireless communication environments (IEEE 802.11ax/be). The proposed solution enables automated auditing and enhances the cyber-resilience of infrastructures containing diverse IoT devices. However, the traditional methods described here, remain static regarding the interpretation of new regulatory requirements, in contrast to RAG-based approaches which offer greater semantic flexibility.

Another good example of using RAG to generate context-aware responses is the research paper [15], which presents a specialized AI-

powered legal assistant that helps users navigate through the Indian tax law. The proposed solution based on a hybrid retrieval mechanism and critique-driven generation proved to be quite effective, achieving 96,5 % accuracy and a 95 % F1-score on a unified Indian tax law benchmark. Nevertheless, the main drawbacks in the context of generating security profiles are the lack of a predefined response format and the focus on the legal framework of a specific country.

To address these issues, it is sufficient to integrate a proprietary database with embeddings derived from current Ukrainian regulations and laws related to information security and cybersecurity, as well as to specify a fixed response format in the form of a table or a list of suggested security controls.

To ensure that the generated content is clear and contextually relevant to the materials provided to the model, it is recommended to use RAG (Retrieval-Augmented Generation) technology. Whilst a standard language model relies solely on what it has learnt during training, hybrid RAG models enhance language generation by combining internal knowledge stored in neural weights (parametric) with real-time information retrieved from an external database (non-parametric) [16]. According to the survey [17], RAG not only improves the accuracy of the responses, but also reduces the number of hallucinations and boosts user trust by citing the sources used.

It is worth noting that while launching a proprietary AI agent is a resource-intensive and labor-intensive process, this approach is a priority from an information security standpoint, as data provided to publicly available versions of the model is not protected against being reused to train the model without the organization's consent [18].

The revealed challenges lead to problem formulation of this work.

## 2. Problem formulation

The transition from a baseline profile to a system-specific target profile is not a straightforward mechanical mapping. The core problem addressed in this paper is therefore defined as follows: how to support the derivation of a draft Target Security Profile from a baseline profile and system-specific context using an LLM-assisted methodology that remains traceable, regulation-aware, and human-supervised.

Let $S$ denote the structured description of a specific information and communication system, including its purpose, architecture, data categories, user roles, external interfaces, and operational constraints (system passport describing the protected information system). Let $B = \{b_1, b_2, \ldots, b_n\}$ denote the set of controls included in the selected baseline profile, together with their minimum required parameters. Let $R$ denote the corpus of relevant regulatory knowledge extracted from ND TPI 3.6-006-24, including control descriptions, implementation guidance, related controls, and enhancement options. The task is to construct a set of draft decisions $T = \{t_1, t_2, \ldots, t_m\}$, where each $t_i$ corresponds to one baseline control $b_i$ of TSP.

Output data should be a structured draft Target Security Profile containing draft decisions for each baseline control, including proposed parameter refinements, enhancements, compensating controls if needed, rationale, and verification artifacts.

## 3. Target Security Profile derivation method

### 3.1. The main concepts

Necessary constraints of the method are:
1. The method must rely only on the provided system and regulatory context;
2. The method must preserve traceability between proposed decisions and normative sources;
3. The method must explicitly indicate cases where the available information is insufficient;
4. The final approval of the target profile must remain with the human expert.

From this perspective, the main methodological challenge is to design a formalized scenario, and workflow in which an LLM, supported by retrieval-augmented generation, can consistently transform regulatory and system-specific inputs into a control-by-control draft profile without producing unsupported assumptions.

## 3.2. Formalized Scenario for Target Security Profile Derivation

Let the derivation of a Target Security Profile be modeled as a constrained decision-support process

$$P: (S, Q, B, R) \to T,$$

where $S$ is the system model derived from the system passport; $Q$ is the set of security requirements for the specific information and communication system; $B = \{b_1, b_2, \ldots, b_n\}$ is the selected baseline security profile; $R$ is the regulatory knowledge base extracted from ND TPI 3.6-006-24; $T = \{t_1, t_2, \ldots, t_m\}$, is the resulting Target Security Profile.

In this setting, each baseline control $b_i \in B$ is represented as

$$b_i = \langle id_i, name_i, p_i^{min}, e_i, rel_i \rangle,$$

where $id_i$ is the control identifier, $name_i$ is the control name, $p_i^{min}$ is the set of minimum baseline parameters, $e_i$ is the set of possible enhancements, and $rel_i$ is the set of related controls or dependencies defined in the regulatory source. ND TPI 3.6-006-24 explicitly treats the baseline profile as the minimum mandatory starting point and defines controls together with their implementation characteristics, including parameterized controls and optional enhancements.

### 3.2.1. System context formalization

The system passport is transformed into a structured model

$$S = \langle A, D, U, Adm, I, Ext, C, M, F \rangle,$$

where $A$ is the set of architectural components; $D$ is the set of data categories processed by the system; $U$ is the set of user roles; $Adm$ is the set of administrator roles; $I$ is the set of internal and external integrations; $Ext$ is the set of exposed external services or interfaces; $C$ is the set of deployment characteristics, including cloud, mobile, and remote access features; $M$ is the set of mission or business functions supported by the system; $F$ is the set of critical failure points or critical assets.

This stage corresponds to the collection of the initial system information required for context-aware profiling.

### 3.2.2. Security requirements extraction

The set of security requirements is defined as

$$Q = \{q_1, q_2, \ldots, q_k\},$$

where each requirement $q_j$ belongs to one or more protection objectives:

$$q_j \in \{\text{Confidentiality, Integrity,} \\ \text{Availability, Accountability,} \\ \text{Privacy, Continuity}\}.$$

A requirement mapping function is then introduced:

$$\phi_Q: S \to Q,$$

which derives the relevant security requirements from the system context. This is consistent with the regulatory logic that security controls are selected as instruments for satisfying security requirements.

### 3.2.3. Security categorization and baseline selection

Let the security category of the system be defined as

$$cat(S) \in \{\text{Low, Moderate, High}\}.$$

Then the baseline selection function is

$$\phi_B: cat(S) \to B.$$

Thus, the baseline profile is selected as

$$B = \phi_B(cat(S)).$$

This reflects the rule that the baseline profile is chosen according to the system's category and serves as the minimum initial set of controls.

### 3.2.4. Baseline expansion into a working matrix

The selected baseline profile is expanded into a working decision matrix $W = \{w_1, w_2, \ldots, w_n\}$, where each row corresponds to one control from the baseline profile:

$$w_i = \langle b_i, p_i^{min}, ctx_i, req_i, risk_i \rangle.$$

Here $ctx_i$ is the subset of system context relevant to control $b_i$; $req_i \subseteq Q$ is the subset of security requirements addressed by $b_i$; $risk_i$ is the set of risks mitigated by $b_i$.

The baseline profile example already shows that controls are not merely listed as identifiers, but instantiated as parameterized actions with minimum required settings, such as timelines, review frequencies, and reporting thresholds. For example, AC-2 contains minimum timing parameters, AC-7 includes organization-defined thresholds and responses for failed logon attempts, and IR-3 and IR-6 specify minimum frequencies and reporting times.

### 3.2.5. Control-level contextual analysis

For each baseline control $b_i$, the system computes a relevance score

$$\rho_i = Rel(b_i, S), \rho_i \in [0,1],$$

where $Rel(\cdot)$ is a relevance function that estimates how strongly the control applies to the specific system context.

At the same time, adequacy of minimum baseline parameters is assessed by

$$\alpha_i = Adeq(p_i^{min}, S, Q, risk_i),$$

where

$$\alpha_i \in \{sufficient, insufficient, unknown\}.$$

The analysis also considers dependencies and environment-specific factors, such as privileged access, remote access, public-facing services, external system interfaces, and cloud deployment. Those factors are especially relevant because the ND TPI catalog includes dedicated control families for access control, incident response, continuity, identification and authentication, perimeter protection, and system interconnection.

### 3.2.6. Control decision function

For each control $b_i$ a decision is assigned from the following set:

$$d_i \in D = \{Keep, Refine, Enhance, Add, Compensate\}.$$

The decision function is defined as

$$d_i = \delta(b_i, S, Q, R, risk_i, \alpha_i).$$

Its logic can be formalized as:

$$d_i = \begin{cases} Keep, if \; \rho_i \geq \theta_r \wedge \alpha_i = Suf; \\ Refine, (\rho_i \geq \theta_r) \wedge (\alpha_i = InSuf) \wedge (b_i \in App); \\ Enhance, if \; (\rho i \geq \theta r) \wedge (risk_i > \theta_{risk}) \wedge \exists e_i; \\ Compensate, if \; b_i \in Infeasible; \\ Add, if \; \exists c \notin B, c \; is \; required \; by \; S, Q, risk, \end{cases}$$

where $\theta_r$ is the control relevance threshold and $\theta_{risk}$ is the risk escalation threshold, $Suf$ means sufficient, $App$ means applicable. This reflects the regulatory logic that controls may require parameterization, strengthening through enhancements, or supplementation when the operational environment demands stronger protection.

### 3.2.7. Parameter instantiation for the Target Security Profile

Each resulting target control record is represented as

$$t_i = \langle id_i, d_i, p_i^{tar}, enh_i, comp_i, r_i, owner_i, art_i \rangle,$$

where $p_i^{tar}$ is the final parameter set proposed for the Target Security Profile; $enh_i$ is the selected set of enhancements, if any; $comp_i$ is the compensating control, if any; $r_i$ is the rationale for the decision; $owner_i$ is the responsible role; $art_i$ is the verification artifact.

The parameter instantiation function is

$$p_i^{tar} = \psi(p_i^{min}, S, Q, risk_i, d_i).$$

Thus, a generic control is transformed into a concrete, auditable rule.

*Example.* AC-2⇒Deactivate account within 8 hours after termination, AC-2⇒Deactivate account within 8 hours after termination, IR-6⇒Report suspicious incidents within 2 hours, IR-6⇒Report suspicious incidents within 2 hours, IR-3⇒Test incident response capability at least annually, IR-3⇒Test incident response capability at least annually. This parameterized form is directly supported by the baseline profile example.

### 3.2.8. Addition of enhancements and extra controls

If the system-specific risk level exceeds the typical assumptions of the selected baseline profile, the Target Security Profile may include additional enhancements or extra controls:

$$T^+ = T \cup E^+ \cup C^+,$$

where $E^+$ is the set of added enhancements; $C^+$ is the set of additional controls not originally included in $B$.

The inclusion criterion can be expressed as

$$e \in E^+ \text{ or } c \in C^+ \iff RiskGap(S, B) > 0.$$

It means, if the baseline profile leaves residual risk that is unacceptable in the specific environment, the profile must be strengthened.

### 3.2.9. Compensation modeling

If a control cannot be implemented directly, the model introduces a compensating control relation

$$\kappa: b_i \to c_i^{comp},$$

where $c_i^{comp}$ is a compensating measure satisfying the expression:

$$Mit(c_i^{comp}) \approx Mit(b_i),$$

that is, the mitigation effect of the compensating control is approximately equivalent, or at least acceptable relative to the intended effect of the original control. The compensation record may be written as

$$c_i^{comp} =$$
$$= \langle reason_i, residualRisk_i, c_i^{comp}, justification_i \rangle.$$

### 3.2.10. Assignment of responsibility and evidence

For each target control $t_i$, the model assigns: $owner_i = \omega(t_i), art_i = \eta(t_i)$, where: $\omega$ maps the control to the responsible organizational role; $\eta$ maps the control to one or more implementation evidence artifacts, such as policy documents, logs, configuration snapshots, incident records, training reports, or review protocols. This step is necessary because the target profile must be implementable and verifiable, not merely descriptive.

### 3.2.11. Expert review and approval

The generated draft Target Security Profile should be then reviewed by a human expert group. It should obtain formal approval by the system owner or authorized decision-maker. Therefore, the complete process can be summarized as

$$T = \gamma(P(S, Q, B, R)),$$

where $\gamma$ is approval process function. This preserves from incorrect LLM decisions by the use of the "human-in-the-loop" principle.

## 4. RAG-oriented workflow architecture

Let us define define the stages of development of an LLM-based advisor.
1. *Data Preparation.* Usually consists of the following steps [17]:
- *Data ingestion & Cleaning*: Removing formatting, unnecessary spaces, advertisements, headers and footers that do not contribute to the content if present. Converting all documents into a single text format suitable for further processing.
- *Chunking*: Splitting large documents into smaller parts to improve search results.

The most common approaches are: character splitting (simply dividing initial document into N-charachter sized parts, often with a certain number of overlapping characters), sentence-level chunking (takes the structure of the text into account and does not split individual sentences) and optimized chunking (maximizes RAG retrieval precision by segmenting documents into small, context-aware pieces while preserving the semantic integrity of tables, images, and hierarchical headings) [19].

- *Embedding*: Using a special embedding model (e.g. [20]), the text is converted into vectors (multidimensional coordinates) which represent its semantic meaning. Thus texts with similar content will have close coordinates in vector space.
- *Indexing & Metadata enrichment*: Adding additional information to chunks, enabling hybrid search (for example, filtering results by date or a specific section of the document).
- *Storage*. Choosing a suitable vector database (e.g. [21]), indexing and uploading the prepared embeddings.

2. *LLM Setup*. Selecting a model (GPT-4o, Llama 3, etc.) and configuring the parameters.
3. *Agent Logic and Prompt Engineering*. Designing system prompts, defining constraints and response formats, and implementing data sufficiency analysis logic.
4. *RAG Integration*. Building the context retriever function, configuring the combining of prompts with the retrieved context.
5. *Agent Deployment*. Deploying the agent in the chosen environment (cloud or on-premise) and configuring monitoring for accuracy and hallucinations.

Once the LLM-based advisor has been implemented, a clear workflow needs to be established. BPMN (Business Process Modelling Notation) was used to visualise the workflow, as it provides a clear, shared standard that works for everyone - from the initial process designers and the technical teams who implement them to the business leaders who monitor the results [22]. The proposed workflow architecture model is detailed in Figure 2.

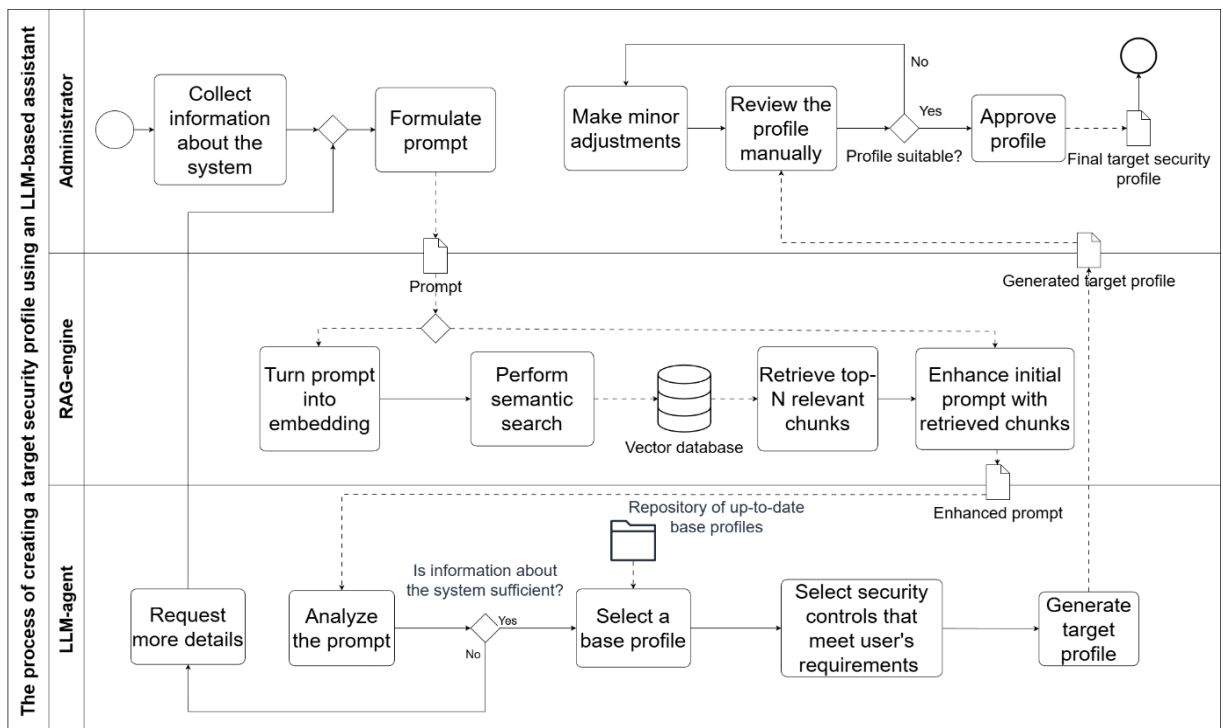

**Figure 2**: Target Security Profile generation workflow

It starts with the administrator gathering system information and formulating a prompt. Information about the system should be as comprehensive as possible so that the model can extract enough context from the prompt. Such information may include the type of data being processed, the system class, and details about the organization and its structure. The RAG-engine then processes this prompt by turning it into an embedding and searching a vector database for relevant data. It uses these findings to create an enhanced prompt. Next, the LLM-agent takes this enhanced prompt, analyzes it, and selects a base profile from a repository. It then chooses specific security controls and generates a target profile. Finally, the administrator reviews the profile. If it does not meet the requirements, they make adjustments and review it once again until they are satisfied, then approve the result as the final target security profile.

An example of a universal prompt for LLM agents is given in an article repository on GitHub [23].

The roles and deliverables during the process of TSP derivation are proposed in Table 1. The results of the profile generation and supportive information is recommended to be gathered for TSP profile are shown in Table 2.

## 5. Case study and results discussion

To verify the adequacy and functionality of the proposed solutions, a system specification (containing synthetic data) was generated; however, this specification fully corresponds to the real-world scenario. The source code for the specification is available on GitHub [23].

It is recommended to create and maintain a comparative table of improvements against the base profile, see the example in Table 2. As a result of applying this methodology, three target reports were obtained that exhibit a number of improved characteristics for the system. The LLMs agents used were Claude Sonnet 4.6 Extended [23], Gemini 3.1 Pro[24], and ChatGPT Thinking 5.4 with Extended effort [25].

The results were evaluated by experts in the field of cybersecurity compliance, and were found satisfiable. For qualitative parameters the scale {Low, Below Medium, Medium, Above Medium, High} were used. The results of validation are shown in Table 3.

**Table 1**
The general roles and process deliverables according to methodology

| Stage | Input | Executor | Results |
|---|---|---|---|
| 1 | Description of the ICS, data type, architecture, users, risks | Employee/Work group | System specification |
| 2 | System specification + ND TPI + base profile | RAG engine | Relevant sections on measures, parameters and amplifications |
| 3 | Context + retrieved chunks | LLM | Draft decisions for each control |
| 4 | Draft solutions | Expert/Administrator | Verified solutions |

**Table 2**
Improved parameters for target profile

| Field | AC-2 | AC-7 | IR-6 | IR-3 | AC-17 | AC-19(5) |
|---|---|---|---|---|---|---|
| Name | Account Management | Unsuccessful Logon Attempts | Incident Reporting | Incident Response Testing | Remote Access | Mobile Device Access Control |
| Baseline parameters | End of user working day | Notify responsible administrator | 2 hours | At least annually | Authorize and control each remote access type | All mobile devices processing organizational data |

| Field | AC-2 | AC-7 | IR-6 | IR-3 | AC-17 | AC-19(5) |
|---|---|---|---|---|---|---|
| TSP parameters | To be refined for the ICS | To be refined for the ICS | Report within 2 hours | Test at least annually | Restrict and control remote access | Encrypt and control relevant mobile devices |
| Enhancements | AC-2(5) | - | - | - | AC-17(3), AC-17(4) | - |
| Notes | Session/account handling parameterized | Failed logon response includes admin notification | Explicit response time | Explicit testing frequency | Includes managed control points | Applies to all relevant organizational devices |

**Table 3**
Improved parameters for target profile

| | Regulatory relevance | Adequacy of decision class | Adequacy of parameters | Quality of rationale | Need for correction | Controls required only minor edits, % | Decisions matched expert judgment directly, % | Hallucinations occurred when system passport lacked details, % |
|---|---|---|---|---|---|---|---|---|
| Claude (Sonnet 4.6 Extended) | High | High | Above Medium | High | Below Medium | 15 | 75 | 10 |
| Gemini (3.1 Pro) | Above Medium | Above Medium | Medium | High | Medium | 25 | 60 | 43 |
| ChatGPT (Thinking 5.4, Extended effort) | High | High | Above Medium | Above Medium | Below Medium | 28 | 65 | 25 |

**Results discussion**.

The experimental results cover the scenario steps described in Sections 3.2.1–3.2.8, 3.2.10, and 3.2.11. Section 3.2.9, Compensation modeling, requires more detailed investigation using examples of real systems. The presented outputs include this field at the structural level, but do not demonstrate a non-trivial infeasible-control case that would require an actual compensating measure.

The results indicate that Claude Sonnet 4.6 Extended achieved the best overall performance, generating the most complete report in under four minutes from a single prompt. ChatGPT Thinking 5.4, with the additional Extended effort setting, provided a detailed 60-page report in 17 minutes and 41 seconds, strictly adhering to the format and adding almost no custom text formatting. Compared to Claude, more controls were flagged as requiring clarification from the user, and significantly fewer controls received recommendations regarding refinement or enhancement.

Gemini 3.1 Pro did not generate a comprehensive report; instead, it provided an analysis in segments of 9–11 controls, followed by a summarized overview.

The experimental results show that the proposed methodology is feasible not only at the conceptual level but also in practical target profile drafting. Quantitative validation confirms this: for the top-performing model (Claude Sonnet 4.6), 15% of generated controls required only minor expert edits, and 75% of decisions matched expert judgment directly.

At the same time, the experiments revealed substantial inter-model differences. Claude produced the most operationally useful output, generating a detailed 60-page report in under 4 minutes. ChatGPT Thinking 5.4 demonstrated a more conservative behavior, frequently flagging controls for clarification, which aligns with the methodology's constraint to avoid unsupported assumptions. In contrast, Gemini 3.1 Pro provided less consistent results, delivering analysis in segments rather than a comprehensive report.

An important finding is the correlation between input quality and model reliability. Hallucinations occurred in a considerable proportion of cases where the system passport was incomplete or lacked essential details (noted in 43% of such cases for Gemini). This reinforces the conclusion that the LLM-based advisor serves as a decision-support tool where the human expert remains the final authority in the 'human-in-the-loop' workflow.

A direction for future research is the development of mechanisms to support Compensation modeling, which was defined in this paper as part of the scenario for creating a universal Target Security Profile (TSP).

## 6. Acknowledgements

The authors would like to express their gratitude to Fedir Yalbuhan, Electronics Engineer of the Department of Technical Information Security at UKRINFORMSYSTEMS LLC, for his expert guidance and leadership in managing the expert evaluation process.

The authors also acknowledge the use of OpenAI's ChatGPT for grammar refinement and language clarity during manuscript preparation

## 7. Conclusions

The proposed methodology for automated security profiling, based on RAG-enhanced Large Language Models, demonstrates that the complex process of selecting security controls under the ND TPI 3.6-006-24 can be transformed into a structured and transparent operational cycle. The primary result of this work is the development of a formalized workflow for deriving a Target Security Profile that achieves high decision accuracy, with up to 80% of AI-generated decisions directly matching expert judgment. The practical significance of this research lies in the implementation of a human-in-the-loop architecture, where the AI manages the analytical burden of reviewing hundreds of parameters while leaving final approval and accountability to the human expert, thereby minimizing risks associated with human error and manual complexity. These findings open significant perspectives for creating universal intelligent decision-support systems that allow organizations to dynamically adapt their defenses to high-intensity hybrid threats, scale compliance with evolving regulatory norms without a proportional increase in specialized staff, and integrate automated security management directly into the lifecycle of critical information systems.